\documentclass[a4paper]{jpconf}
\usepackage{anyfontsize}
\usepackage[draft]{todonotes} 
\usepackage{graphicx}
\usepackage{amsmath}
\begin{document}
\renewcommand{\arraystretch}{1.5}
\title{Quantum statistics in Bohmian trajectory gravity}

\author{T C Andersen}

\address{NSCIR - 046516 Meaford, Ontario N4L 1W7, Canada}

\ead{tandersen@nscir.ca}

\begin{abstract}
The recent experimental proposals by Bose et al. and Marletto et al. (BMV) outline a way to test for the quantum nature of gravity by measuring gravitationally induced differential phase accumulation over the superposed paths of two $\sim 10^{-14}kg$ masses. These authors outline the expected outcome of these experiments for semi-classical, quantum gravity and collapse models. It is found that both semi-classical and collapse models predict a lack of entanglement in the experimental results. This work predicts the outcome of the BMV experiment in Bohmian trajectory gravity - where classical gravity is assumed to couple to the particle configuration in each Bohmian path, as opposed to semi-classical gravity where gravity couples to the expectation value of the wave function, or of quantized gravity, where the gravitational field is itself in a quantum superposition. In the case of the BMV experiment, Bohmian trajectory gravity predicts that there will quantum entanglement. This is surprising as the gravitational field is treated classically. A discussion of how Bohmian trajectory gravity can induce quantum entanglement for a non superposed gravitational field is put forward. 
\end{abstract}

\section{Introduction}
In papers published in 2017, Bose et al.\cite{Bose}, and Marleto and Vedral\cite{Marletto2017PRL} (BMV) describe an experimental proposal to test for quantum gravity . Christodoulou and Rovelli\cite{Christodoulou} summarize  these papers and clarify the quantum - gravity interaction in these proposals. Christodoulou and Rovelli develop a simple common language to describe the experiment, using an assumption that while perhaps not possible in an actual experiment, simplifies discussion and understanding of the effect. The notation and terminology of the Christodoulou and Rovelli paper will be used here. 

We will first review the BMV experimental setup, defining base states and terminology. Then the quantized gravity, semi-classical and collapse predictions will be recovered. 

A section follows on how Bohmian trajectory gravity would be interpreted for the BMV experiment. The experimental predictions for Bohmian quantum gravity are then explored. 

\section{The BMV experiment}

The in the BMV experimental proposal, Bose\cite{Bose} and Marletto and Vedral\cite{Marletto2017PRL} propose, in different experimental setups to measure the same effect, namely the expected behavoir of quantum systems when particles interact only via mutual graviational attraction. Christodoulou and Rovelli\cite{Christodoulou} point out that the effect can be understood as the time dilation effects of one mass on the other while each particle is in a spatial superposition. Although the time dilation effect is tiny, the experimental effect is amplified as the experimenters can take advantage of the extremely high Compton frequency of the massive (in the quantum particle sense) particle. The experimenters only need to measure the \textit{difference} in phase accumulation along different paths, and once that phase difference is $\sim \pi$ the experimenters will detect that the state has become quantum entangled. 
For details on how a heavy quantum particle with spin ready for a Stern-Gerlach device can be made and kept in a superposition, see Bose et al.\cite{Bose}  

\begin{figure}[ht]\label{bmvfig}
\includegraphics[width=19pc]{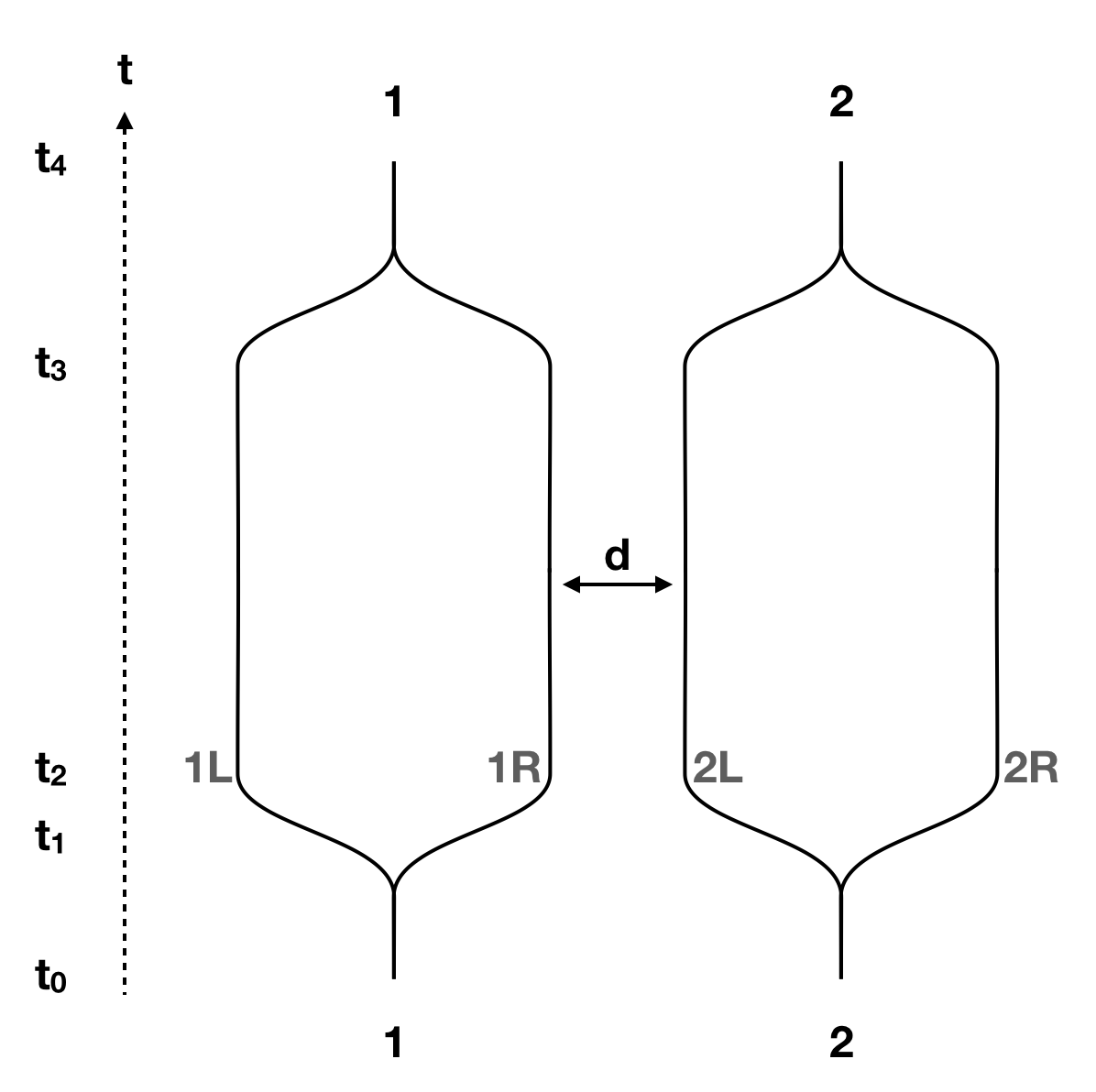}\hspace{2pc}
\begin{minipage}[b]{14pc}\caption{\label{label5}Particles 1 and 2 start off in the spin up $ |u\rangle$ (defined as +z) direction, and are spatially split into superposed $ |L\rangle + |R\rangle$ states by a Stern-Gerlach mechanism. The particles fly through the experiment at the same time, and then through a second Stern-Gerlach where they are brought back into up/down state. If there is negligible gravitational interaction, (for example with particles such as silver atoms) the final state will be $|u\rangle \otimes |u\rangle$ (times an overall phase factor) and the particles are not entangled.}
\end{minipage}
\end{figure}

\section{Measuring quantum entanglement}
We wish to determine the quantum entanglement of the final state in the quantum gravity models considered in this paper. In some cases such as semi-classical and quantized gravity case the final state is pure, while in others a mixed state is the result. For example a mixed state will result from environmental collapse. Quantum entanglement cannot be measured using partial traces and von Neumann entropy on mixed states.\cite{Vedral1997}. Rather, a well chosen entanglement witness is a good measure.  We will in this paper use a witness similar to that of Bose et al.\cite{Bose}. Witness:

\begin{eqnarray}\label{witness}
{\cal W}=\langle \sigma_x^{(1)} \otimes \sigma_z^{(2)} \rangle + \langle \sigma_y^{(1)} \otimes \sigma_y^{(2)} \rangle
\end{eqnarray}

which shows entanglement when
 
\begin{eqnarray}
{\cal W} \leq 0 .
\end{eqnarray}

We use the convention
\begin{eqnarray}
u = \begin{bmatrix}
1\\
0
\end{bmatrix}, & 
d = \begin{bmatrix}
0\\
1
\end{bmatrix} & 
\end{eqnarray}

\begin{eqnarray}
\sigma_x = \begin{bmatrix}
0 & 1\\
1 & 0
\end{bmatrix} & 
\sigma_y = \begin{bmatrix}
0 & -i\\
i & 0
\end{bmatrix} & 
\sigma_z = \begin{bmatrix}
1 & 0\\
0 & -1
\end{bmatrix}.
\end{eqnarray}

So that the operator $W$ of the witness $\cal W$ is 
\begin{eqnarray}
W = \sigma_x^{(1)} \otimes \sigma_z^{(2)} + \sigma_y^{(1)} \otimes \sigma_y^{(2)} = \begin{bmatrix}
\phantom{I}\ 0 & \ 0 & \ 1 & \ \llap{$-$}1\\
\phantom{I}\ 0 & \ 0 & \ 1 & \ \llap{$-$}1\\
\phantom{I}\ 1 & \ 1 & \ 0 & \ 0\\
\phantom{I}\ \llap{$-$}1 & \ \llap{$-$}1 & \ 0 & \ 0
\end{bmatrix}.
\end{eqnarray}

See for instance chapter 2 of \textit{Entanglement and decoherence: foundations and modern trends}\cite{buchleitner2008entanglement}. The witness $\cal W$ (\ref{witness}) is thus in the range [-2, 2].

The authors also found the online entaglement calculator\cite{Technion-IsraelInstituteofTechnologyHaifa} useful. The mathematics package Maple was  employed to handle density matrix and witness calculations. 

In this paper we will always express the final state with a density matrix, even if the final state is pure. This simplifies the analysis. The density matrix will be given in the $|u\rangle, |d\rangle$ basis.

With a density matrix $\rho$, the expression to evaluate the entanglement witness is simply:
\begin{eqnarray}
{\cal W}(\rho) = Tr(\rho W).
\end{eqnarray}

\section{Quantized gravity prediction}
In quantized gravity approaches to quantum gravity, the gravitational field can/will be in a superposition. Theories like canonical, Wheeler-Dewitt, and loop quantum gravity all predict that the gravitational field will be in a superposition, and that at low energy, (i.e. perturbatively) will behave the same way, as a quantized gravity. Thus the BMV experiment will not be able to distinguish between these quantized gravity models. Rather the expectation is that the results of the BMV experiment will show that gravity is quantized.

The initial state is, with $u$ representing a $+z$ spin direction,  

\begin{eqnarray}
|\Psi_{t_0}\rangle&=& |u_1\rangle \otimes  |u_2\rangle \otimes  |g\rangle .
\end{eqnarray}

Following Christodoulou and Rovelli \cite{Christodoulou}: 
\begin{quotation}
We assume that $|g\rangle$ belongs to a Hilbert space that contains semiclassical states that approximate classical geometries $g$, \textit{but also linear superposition of these}. This is the key property needed to derive the BMV effect. 
\end{quotation}

Each particle is passed through a typical Stern-Gerlach device, where L and R represent spins in the $\pm x$ direction.
 \begin{eqnarray}
|u_1\rangle=\frac{|L_1\rangle+|R_1\rangle}{\sqrt{2}}\ \ , \ \ 
|u_2\rangle=\frac{|L_2\rangle+|R_2\rangle}{\sqrt{2}} .
\end{eqnarray}

So that at $t_1$ we have 

\begin{eqnarray}
|\Psi_{t_1}\rangle&=&\frac12\,\Big(|L_1\rangle+|R_1\rangle)  \otimes   (|L_2\rangle+|R_2\rangle\Big) \otimes  |g\rangle
 \\
&=&\frac12\, \Big(|LL\rangle+|RR\rangle + |LR\rangle+|RL\rangle\Big) \otimes  |g\rangle. 
\end{eqnarray}

At this point the gravitational field is felt differently by all 4 branches, and the state quickly becomes

\begin{eqnarray}
|\Psi_{t_2}\rangle&=&\frac12\, \Big( |LL\rangle\otimes |g_{d_{LL}}\rangle+|RR\rangle \otimes |g_{d_{RR}}\rangle
\nonumber \\ && \ \ \  
+ |LR\rangle \otimes |g_{d_{LR}}\rangle+|RL\rangle \otimes  |g_{d_{RL}}\rangle\Big).
\end{eqnarray}

Christodoulou and Rovelli \cite{Christodoulou} assume the experiment is set up such that only $|g_{d_{RL}}\rangle$ is of physical consequence, the other gravitational states being too weak to affect their respective branches. 

The state evolves over time to $t_3$ where the additional phase $e^{i\frac{Gm^2 t}{\hbar d}}$ appears on the closest approach branch, while the huge overall phase evolution on all 4 branches is $e^{-i\frac{mc^2 t}{\hbar}}$, where $t = t_3 - t_2$ is the time that the particles spend gravitationally interacting.
\begin{eqnarray}
|\Psi_{t_3}\rangle&=&\frac{e^{-i\frac{mc^2 t}{\hbar}}}{2}\, \Big(|LL\,g_{d_{LL}}\rangle+|RR\,g_{d_{RR}}\rangle \\
&& \nonumber +|LR\,g_{d_{LR}}\rangle+e^{i\frac{Gm^2 t}{\hbar d}} |RL\,g_{d_{RL}}\rangle\Big). 
\label{stato2}
\end{eqnarray}

Dropping the overall phase factor as it does not change the physical state, and at $t_4$ recombining the branches with another Stern-Gerlach device, the gravitational field is again in one state, with $t = t_3 - t_2$ chosen such that $e^{i\frac{Gm^2 t}{\hbar d}} = -1$, the state becomes:

\begin{eqnarray}\label{qgfinal}
|\Psi_{t_4}\rangle_{qg}&=&\frac12\, \Big(|LL\rangle+|RR\rangle + |LR\rangle- |RL\rangle\Big)\otimes |g\rangle.    
\end{eqnarray}

Christodoulou and Rovelli\cite{Christodoulou} trace out the gravity and final state of one particle, arriving at the conclusion that the state (\ref{qgfinal}) is a maximally entangled pure state. 
 
Here we arrive at the same conclusion by first calculating the density matrix for the final state (\ref{qgfinal}): 
\begin{eqnarray}\label{qgrho}
\rho_{qg}=\frac14
\begin{bmatrix}
\phantom{I}\ 1 & \ 1 & \ \llap{$-$}1 & \ 1 \\
\phantom{I}\ 1 & \ 1 & \ \llap{$-$}1 & \ 1 \\
\phantom{I}\ \llap{$-$}1 & \ \llap{$-$}1 & \ 1 & \ \llap{$-$}1 \\
\phantom{I}\ 1 & \ 1 & \ \llap{$-$}1 & \ 1
\end{bmatrix}   .  
\end{eqnarray}

Calculating the witness (\ref{witness}) on the density matrix $\rho_{qg}$ shows:
\begin{eqnarray}
{\cal W}(\rho_{qg}) = Tr(\rho_{qg} W) = -2 & (quantum\ gravity).
\end{eqnarray}

\section{Semi-classical gravity prediction}
While semi-classical gravity is mostly thought of as an aid to calculate the effects of gravity on quantum systems, there are some researchers\cite{Grobart2017} who look at semi-classical gravity as a real possibility. In semi-classical gravity the gravitational field of a particle in a quantum state is taken to be the expectation value of the position of the particle. 

It's worthwhile to recall the physical model of semi-classical gravity:
\begin{equation}\label{semiclassEE} R_{\mu\nu}-\dfrac{1}{2}g_{\mu\nu}R=\dfrac{8\pi G}{c^4}\langle\Psi|T_{\mu\nu}|\Psi\rangle .
\end{equation}

We see that Einstein's gravity connects  to the expectation value of the matter-energy configuration $T$. Thus in this scenario, gravity is somehow able to violate the measurement postulates of quantum mechanics, as the gravitational field always knows the zeroth moment of the position operator on an arbitrary wave function. 

In the case of the BMV experiment, in the configuration considered here, with the distance $d$ in Figure \ref{bmvfig} small in relation to the $L - R$ split distance on each particle, the experimental outcome is not changed from the 'light' particle prediction, and the final state is the same as the intial state:

\begin{eqnarray}\label{scfinal}
|\Psi_{t_4}\rangle_{sc}&=&\frac12\, \Big(|LL\rangle+|RR\rangle + |LR\rangle+ |RL\rangle\Big),    
\end{eqnarray}

which is of course separable, and can be written as 
\begin{eqnarray}\label{scfinal2}
|\Psi_{t_4}\rangle_{sc}&=|u\rangle \otimes |u\rangle,    
\end{eqnarray}

\begin{eqnarray}\label{scrho}
\rho_{sc}=
\begin{bmatrix}
1 & 0 & 0 & 0 \\
0 & 0 & 0 & 0 \\
0 & 0 & 0 & 0 \\
0 & 0 & 0 & 0
\end{bmatrix} .     
\end{eqnarray}

As a check, running the witness

\begin{eqnarray}
{\cal W}(\rho_{sc}) = Tr(\rho_{sc}W) = 0 & (semi-classical\ gravity).
\end{eqnarray}

\section{Collapsed state prediction}
The BMV experiment is not simple to build or execute, and it is entirely possible that the superposed states will collapse into L or R for each of the particles irrespective of the other particle. There are also collapse models where particles as massive as that used in the proposed BMV experiment will collapse the wave function.\cite{Kafri2014r} In these cases the final state will be a mixed state where we know the source of mixing and result will be one of in equal mixture:

\begin{eqnarray}\label{colfinal}
|LL\rangle, |RR\rangle,|LR\rangle, |RL\rangle,    
\end{eqnarray}

represented with a density matrix,

\begin{eqnarray}\label{colrho}
\rho_{col}=\frac14
\begin{bmatrix}
1 & 0 & 0 & 0 \\
0 & 1 & 0 & 0 \\
0 & 0 & 1 & 0 \\
0 & 0 & 0 & 1
\end{bmatrix}     .
\end{eqnarray}

With the witness 
\begin{eqnarray}
{\cal W}(\rho_{col}) = Tr(\rho_{col}W) = 0 & (collapse).
\end{eqnarray}

\section{Bohmian trajectory prediction}
de Broglie Bohm mechanics, as begun by Louis de Broglie and formalized by David Bohm\cite{Bohm1952} uses non local hidden variables to create an interpretation of quantum mechanics where particles have trajectories, with an ensemble of trajectories traversing the entire wave function.

 If one were to use a Bohmian description of quantized gravity, then the results would be identical to those of the quantized gravity predictions and equation (\ref{qgfinal}) would describe the final state. 

Instead, what is assumed for Bohmian trajectory gravity is that as Struyve writes:\cite{STRUYVE2017}
\begin{quotation}
	Bohmian mechanics solves the measurement problem by introducing an actual configuration (particle positions in the non-relativistic domain, particle positions or fields in the relativistic domain) that evolves under the influence of the wave function. According to this approach, instead of coupling classical gravity to the wave function, it is natural to couple it to the actual matter configuration.
\end{quotation}

In equation form, comparing to the semi-classical form \ref{semiclassEE} above, we have:
\begin{equation}
R_{\mu\nu}-\dfrac{1}{2}g_{\mu\nu}R = 8\pi G T_{\mu \nu}(\varphi_{B},g) \,.
\label{bohmtrajEE}
\end{equation}

Where $\varphi_{B}$ is the trajectory of one member of the ensemble of Bohmian trajectories, and $T_{\mu \nu}(\varphi_{B},g)$ is the stress energy tensor as determined by the specific trajectory in combination with perhaps other (e.g. earth's gravity) sources. A simple interpretation of a quantum - gravity coupling such as (\ref{bohmtrajEE}) has theoretical problems such as energy violation, which it shares with other similar approaches.\cite{Pinto-Neto2018}

In order to interpret what happens in the BMV experiment assuming Bohmian trajectory gravity governs the quantum - gravity coupling consider
\begin{description}
	\item [Superposition:] since gravity couples directly to each particle position, gravity is not in a superposition, instead the configuration of the gravitational field changes for each run of the experiment, with 4 possible configurations of particle trajectories. 
	\item [No interaction:] With the BMV experiment set up as outlined in Figure \ref{bmvfig}, in three of the four configurations there is no substantial gravitational interaction. 
	\item [Interaction:] only when particle 1 takes the right track and particle 2 takes the left track will there be a gravitational interaction. 
	\item [Measurement:] There is no measurement taking place, so there is no collapse. The effect on the quantum state when the particle trajectories take the R and L paths is for the wave function to be altered in the same way as in the quantized gravity solution.  
\end{description}

The end result is that at $t_4$ there are two possible states, we have a mixed state.  

3/4 of the time, there will be no gravitational interaction and the final state will be the same as the starting state at $t_0$:  
\begin{eqnarray}
|\Psi_{t_4}\rangle_{bt}&=&\frac12\, \Big(|LL\rangle+|RR\rangle + |LR\rangle+|RL\rangle\Big) .
\end{eqnarray}
1/4 of the time there is gravitational time dilation and the final state will be the same as that for quantized gravity. The final state is not simply $-|RL\rangle$ as even though 'we know' the particles interacted, the wave function is not collapsed, and interference between the branches will still occur, guiding the particles with the Bohmian guidance equation:  
\begin{eqnarray}
|\Psi_{t_4}\rangle_{bt}&=&\frac12\, \Big(|LL\rangle+|RR\rangle + |LR\rangle-|RL\rangle\Big) .
\end{eqnarray}

In this scenario we have somehow turned a pure state into a mixed state with no measurement (accidental or otherwise). This is to be expected, as gravity is not quantized and the Bohmian trajectory interpretation interacts outside the rules of quantum mechanics. 

Another way of considering this effect: Gravity is linked to a hidden variable, and thus can peer into the quantum wave function. Of course the hidden Bohmian trajectory variable is not hidden from gravity in this case and thus the state can be evolved based on this 'no longer hidden' variable.   

The density matrix for this mixed state is 
\begin{eqnarray}\label{btrho}
\rho_{bt}= \frac{1}{16}
\begin{bmatrix}
\phantom{W}13 && 1  && \llap{$-$}1 && 1 \\
\phantom{W}1 && 1 && \llap{$-$}1 && 1 \\
\phantom{W}\llap{$-$}1 && \llap{$-$}1 && 1 && \llap{$-$}1 \\
\phantom{W}1 && 1 && \llap{$-$}1 && 1
\end{bmatrix}     .
\end{eqnarray}

Evaluating the witness 
\begin{eqnarray}
{\cal W}(\rho_{bt}) = Tr(\rho_{bt}W) = -\frac{1}{2} & (Bohmian\ trajectory).
\end{eqnarray}

The state is entangled. The witness shows a result 1/4 of that of the maximally entangled quantized gravity case, which makes sense - the density matrix in this case it 3/4 mixed with a non entangled state. 

Thus we have created entanglement with a classical single valued field. This seems in violation of the LOCC theorems. But LOCC is built on quantum mechanics - and cannot have any bearing over a scheme such as Bohmian trajectory gravity which violates quantum mechanics. Chitambar et al. in \textit{Everything You Always Wanted to Know About LOCC (But Were Afraid to Ask)}\cite{Chitambar2014} describes LOCC and summarizes:
\begin{quotation}
	Quantum operations implemented in such a manner are known as LOCC (local operations with classical communication), and we can think of LOCC as a special subset of all physically realizable operations on the global system.
\end{quotation}

Other authors\cite{Altamirano2018} discussing the role of gravity with respect to LOCC also share this view. 

\section{Discussion}
The BMV (and related experiments) provide a good probe of the nature of the interaction of gravity and quantum mechanics. The proposed experiment while very challenging to perform, is simple to study theoretically. 

The result prediction for Bohmian trajectory gravity shows entanglement with a witness of $-\frac{1}{2}$, while the quantized gravity approach has an entanglement witness value of $-2$. The semi-classical and collapse models, while having different final states, do not show any entanglement. 

Hall and Reginatto\cite{Hall2018} note that the LOCC theorems assume Koopman-type dynamics, and that there are other classical-quantum interaction schemes. They prove that the result obtained above for semi-classical interactions is generic in that semi-classical mean field models will never create quantum entanglement. They find that there exist other interaction schemes which can create entanglement.

It is interesting to compare the two schemes considered here that connect a classical field of gravity to a quantum particle. The difference between the two interaction schemes is in some sense small. In both the semi-classical and Bohmian trajectory quantum gravity schemes the classical gravitational field must still interact with a quantum entity. In both semi-classical and Bohmian trajectory gravity, the connection lies outside quantum mechanics (non Koopman-style\cite{Hall2018} interactions). This is in contrast to schemes like the KTM\cite{Kafri2014r} and others\cite{Khosla2018} where a way for a classical field to interact with a massive quantum particle is outlined that follows the rules of quantum mechanics. In \cite{Kafri2014r} and \cite{Khosla2018} weak measurements both slowly measure the centre of mass of the particle and also automatically collapse the wave function.

  In some sense semi-classical gravity seems more complicated than Bohmian trajectory gravity, as in semi-classical gravity the gravitational field has to somehow integrate the entire position space of the wave function (a non local entity) in real time (via the Schr{\"o}dinger - Newton equation), in order to continuously use the expectation value as a source for the gravitational field. In Bohmian mechanics, the gravitational field connects directly to an existing 'hidden' particle position, which is conceptually simpler. 
  
  While this paper concerns the BMV experiment in particular, it is not hard to envision applying Bohmian trajectory gravity to other experimental predictions. Applying the method to a BMV experiment where all the branch distances take part is one future extension.  
  
Of course the canonical  viewpoint\cite{Christodoulou} is that the BMV experiment if done without environmental state collapse will side with quantized gravity and show an entangled final state with the entanglement witness approaching -2 as the experiment is performed with increasing accuracy.  

\section*{Acknowledgements}
The author wishes to thank Hans-Thomas Elze and the other organizers and attendees of DICE2018 Spacetime - Matter - Quantum Mechanics Ninth International Workshop in Castiglioncello Italy, for the chance to speak on the matter presented in this paper. 

\section*{References}
\bibliography{library.bib}

\providecommand{\newblock}{}
\begin{thebibliography}{10}
\expandafter\ifx\csname url\endcsname\relax
  \def\url#1{{\tt #1}}\fi
\expandafter\ifx\csname urlprefix\endcsname\relax\def\urlprefix{URL }\fi
\providecommand{\eprint}[2][]{\url{#2}}

\bibitem{Bose}
Bose S, Mazumdar A, Morley G~W, Ulbricht H, Toro{\v{s}} M, Paternostro M,
  Geraci A~A, Barker P~F, Kim M~S and Milburn G 2017 {\em Phys. Rev. Lett.\/}
  {\bf 119} 240401

\bibitem{Marletto2017PRL}
Marletto C and Vedral V 2017 {\em Phys. Rev. Lett.\/} {\bf 119} 240402

\bibitem{Christodoulou}
Christodoulou M and Rovelli C {\em Preprint arXiv:\/} {\bf 1808.05842}

\bibitem{Vedral1997}
Vedral V, Plenio M~B, Rippin M~A and Knight P~L 1997 {\em Phys. Rev. Lett.\/}
  {\bf 78} 2275--2279

\bibitem{buchleitner2008entanglement}
Buchleitner A, Viviescas C and Tiersch M 2008 {\em {Entanglement and
  decoherence: foundations and modern trends}\/} vol 768 (Springer Science {\&}
  Business Media)

\bibitem{Technion-IsraelInstituteofTechnologyHaifa}
Technion {State Separator (V2.0)}
  \urlprefix\url{https://physics.technion.ac.il/stateseparator/}

\bibitem{Grobart2017}
Gro{\ss}ardt A 2017 {\em Proceedings of Science\/} {\bf Corfu Summ}

\bibitem{Kafri2014r}
Kafri D, Taylor J~M and Milburn G~J 2014 {\em New J. Phys.\/} {\bf 16} 1--11

\bibitem{Bohm1952}
Bohm D 1952 {\em Phys. Rev.\/} {\bf 85} 166--179

\bibitem{STRUYVE2017}
Struyve W 2017 {Towards a Novel Approach to Semi-Classical Gravity} {\em The
  Philosophy of Cosmology\/} (Cambridge University Press) pp 356--374 ISBN
  9781316535783

\bibitem{Pinto-Neto2018}
Pinto-Neto N and Struyve W 2018 {\em Preprint arXiv:\/} {\bf 1801.03353}

\bibitem{Chitambar2014}
Chitambar E, Leung D, Manˇcinska L, Ozols M and Winter A 2014 {\em Commun.
  Math. Phys.\/} {\bf 238} 303--326 (\textit{Preprint} \eprint{1210.4583})

\bibitem{Altamirano2018}
Altamirano N, Corona-Ugalde P, Mann R~B and Zych M 2018 {\em Class. Quant.
  Grav.\/} {\bf 35} 145005 (\textit{Preprint} \eprint{arXiv:1612.07735v2})

\bibitem{Hall2018}
Hall M~J and Reginatto M 2018 {\em J. Phys. A: Math. Theor.\/} {\bf 51} 1--18
  (\textit{Preprint} \eprint{1707.07974})

\bibitem{Khosla2018}
Khosla K~E and Nimmrichter S 2018 {\em Preprint arXiv:\/} {\bf 1812.03118}

\end{thebibliography}

\end{document}